\documentclass[a4paper,UKenglish,cleveref, autoref, thm-restate]{lipics-v2021}
\nolinenumbers
\usepackage{amsmath}
\usepackage{amssymb}
\usepackage{wrapfig}
\usepackage{caption}
\usepackage{xcolor}
\usepackage[linesnumbered,boxed]{algorithm2e}
\usepackage{paralist}

\definecolor{blue25}{rgb}{0,0,0.25}
\newcommand{\emphic}[2]{%
     \textcolor{blue25}{%
         \textbf{\emph{#1}}}%
         \index{#2}}
\newcommand{\emphi}[1]{\emphic{#1}{#1}}

\newcommand{\lemlab}[1]{\label{lemma:#1}}
\newcommand{\lemref}[1]{Lemma~\ref{lemma:#1}}

\newcommand{\seclab}[1]{\label{sec:#1}}
\newcommand{\secref}[1]{Section~\ref{sec:#1}}
\newcommand{\thmlab}[1]{\label{theorem:#1}}

\newcommand{\PntSet}{P}

\newcommand{\Mspace}{\mathcal{M}}
\newcommand{\pnt}{p}

\newcommand{\query}{q}
\newcommand{\ball}[1]{\mathsf{B}(#1)}

\renewcommand{\Re}{{\rm I\!\hspace{-0.025em} R}}

\newcommand{\dist}[1]{\mathsf{dist}(#1)}
\newcommand{\Line}{\ell}
\newcommand{\Int}{I}
\newcommand{\IntSet}{\mathcal{I}}
\newcommand{\lept}{a}
\newcommand{\rept}{b}
\newcommand{\CandSet}{\mathcal{C}}
\newcommand{\cpt}{\mathsf{c}}
\newcommand{\cl}[1]{\bar{#1}}

\newcommand{\ropt}{r^*}
\newcommand{\remove}[1]{}

\newcommand{\true}{\mathsf{\mathbf{True}}}
\newcommand{\false}{\mathsf{\mathbf{False}}}

\newcommand{\cardin}[1]{| #1 |}

\SetCommentSty{mycommfont}
\SetKwInput{KwInput}{Input}                % Set the Input
\SetKwInput{KwOutput}{Output}              % set the Output

\title{Red blue \texorpdfstring{$k$}{k}-center clustering with distance constraints} %TODO Please add

\author{Marzieh Eskandari}{Department of Computer Science, Alzahra University, Tehran, Iran \and \url{https://staff.alzahra.ac.ir/eskandari/en/} }{eskandari@alzahra.ac.ir}{https://orcid.org/0000-0002-3302-4338}{}%TODO mandatory, please use full name; only 1 author per \author macro; first two parameters are mandatory, other parameters can be empty. Please provide at least the name of the affiliation and the country. The full address is optional

\author{Bhavika B. Khare}{Department of Computer Science, University of Memphis, Memphis, TN, USA}{bbkhare@memphis.edu}{https://orcid.org/0000-0001-9034-7398}{}

\author{Nirman Kumar}{Department of Computer Science, University of Memphis, Memphis, TN, USA \and \url{https://www.cs.memphis.edu/~nkumar} }{nkumar8@memphis.edu}{https://orcid.org/0000-0002-1825-0097}{}%TODO mandatory, please use full name; only 1 author per \author macro; first two parameters are mandatory, other parameters can be empty. Please provide at least the name of the affiliation and the country. The full address is optional

\authorrunning{M.\,Eskandari and B.\,Khare and N.\,Kumar} %TODO mandatory. First: Use abbreviated first/middle names. Second (only in severe cases): Use first author plus 'et al.'

\Copyright{Marzieh Eskandari and Bhavika B. Khare and Nirman Kumar} %TODO mandatory, please use full first names. LIPIcs license is "CC-BY";  http://creativecommons.org/licenses/by/3.0/

 %TODO mandatory: Please choose ACM 2012 classifications from https://dl.acm.org/ccs/ccs_flat.cfm 

\keywords{Algorithms, Facility Location, Computational Geometry, Clustering} % mandatory; please add comma-separated list of keywords

\EventEditors{John Q. Open and Joan R. Access}
\EventNoEds{2}
\EventLongTitle{42nd Conference on Very Important Topics (CVIT 2016)}
\EventShortTitle{CVIT 2016}
\EventAcronym{CVIT}
\EventYear{2016}
\EventDate{December 24--27, 2016}
\EventLocation{Little Whinging, United Kingdom}
\EventLogo{}
\SeriesVolume{42}
\ArticleNo{23}
\begin{document}
\maketitle

\begin{abstract}
We consider a variant of the $k$-center clustering problem in $\Re^d$, where the centers can be
divided into two subsets, one, the red centers of size $p$, and the other, the blue centers of size $q$, where $p+q=k$, and such that each red center and each blue center must be apart a distance of at least some given $\alpha \geq 0$, with the aim of minimizing the covering radius. We provide a bi-criteria approximation algorithm for the problem and a polynomial time algorithm for the constrained problem where all centers must lie on a given line $\Line$.
\end{abstract}

\section{Introduction}\seclab{intro}
The classic $k$-center problem is to find $k$ balls of minimum radius whose union covers a set $\PntSet$ of $n$ points in a metric space for a given positive integer $k$. This problem provides a simple geometric model for the following facility location problem. We want to
place $k$ facilities (such as supermarkets) to serve the customers in a city. It is natural to assume that the clients go to the facility closest to their home. So, we want to locate $k$ facilities such that the maximum distance between a customer home and the nearest facility is minimized.

The $k$-center problem is known to be NP-hard for Euclidean spaces \cite{megiddosupotwik}.
This paper considers a variant of the $k$-center problem where where the centers can be
divided into two subsets, one, the red centers of size $p$, and the other, the blue centers of size $q$, where $p+q=k$, and such that each red center and each blue center must be apart a distance of at least some given $\alpha \geq 0$, with the aim of minimizing the covering radius. For $\alpha =0$, we get back the $k$-center problem. As a motivating example, suppose that we want to open two types of facilities with the same service (say `Shell' and `BP' gas stations). Each client must have access to at least one of these facilities within the minimum possible distance, but the facilities should be separated from each other to avoid disadvantages of being near competitors (such as crowding,  spying and over-sharing).

Sylvester \cite{sylvester} presented the 1-center problem in 1857, and Megiddo \cite{megiddo} gave a linear time algorithm for solving this problem in 1983, using linear programming. Hwang et al. \cite{hwang} showed that in the plane the $k$-center problem can be solved in $n^{O(\sqrt{k})}$. Agarwal and Procopiuc \cite{agarwal} presented an $n^{O(k^{1-1/d})}$-time algorithm for solving the $k$-center problem in $\Re^d$ and a $(1 + \epsilon)$-approximation algorithm with running time $O(n \log k) + (k/ \epsilon)^{O(k^{1-1/d})}$. Gonzalez \cite{gonzalez} studied approximating the discrete version where the centers must be part of the given point set.

%%%%%%%%%%%%%%%%%%%%%%%%%%%%%%%%%%
%TODO

Some researchers studied constrained versions of the $k$-center problem in which the centers are constrained to a line. Brass et al. \cite{brass} proposed an  $O(n \log^ 2 n)$-time algorithm when the line is fixed in advance. Also, they solved the general case where the line has an arbitrary orientation in $O(n^4 \log^2 n)$ expected time.  Other variations have also been
considered \cite{hurtado,boselangerman,bosetoussaint} for $k=1$. For $k \geq 2$, variants have been studied
as this has applications to placement of base stations in wireless sensor networks \cite{das,roy,shin}. Another variant is the $\alpha$-connected two-center problem, where the goal is to find two balls of minimum radius $r$ whose union covers the points, and the distance of the two centers is at most $2(1 - \alpha)r$, for $0 \leq \alpha \leq 1$. Hwang et al. \cite{hwang} presented an $O(n^2 \log^2 n)$ expected-time algorithm. 

%%%%%%%%%%%%%%%%%%%%%%%%%%%%%%%%%%%%%%%%%%%

Kavand et. al. studied yet another variant of the 2-center problem which they termed as the $(n,1,1,\alpha)$-center \cite{kavand}. The aim was to find two balls each of which covers the whole points, the radius of the bigger one is minimized, and the distance between the two centers is at least $\alpha$. They presented an $O(n\log n)$-time algorithm for this problem, and a linear time algorithm for its constrained version using the farthest point Voronoi diagram. Recently the authors have done some work on this problem generalizing it to the case of two types of centers, red and blue, where every pair of red and blue centers is separated by at least $\alpha$ and balls around all red centers as well as balls around all blue centers cover the point set \cite{eskandari}.
This paper considers a similar looking but different generalization of the $k$-center problem with the $\alpha$-separability assumption, and we denote it by $(n,p \lor q,\alpha)$ problem.
%We explain our choice of notation, particularly the $\land$ sign, in \secref{defs}.
Given a set $\PntSet$ of $n$ points in a metric space and integers $p, q \geq 1$, we want to find $p+q$ balls of two different types, called \emphi{red} and \emphi{blue}, with the minimum radius such that $\PntSet$ is covered by these $p+q$ balls, and the center of each red ball is at least $\alpha$ distant from the center of each blue ball.

%%NK: One motivating example already mentioned and reasons have been explained. Removing for space constraints.
%As mentioned before, in $(n,p \lor q,\alpha)$ problem, there are two types of facilities to locate in a city. These facilities may be two different brands of a same service facility  with the same qualities (such as `Costco's' and `Sam's club'). We need to cover each client with at least one facility within the nearest distance, but businesses need to be careful to not share their trade secrets. Moreover, depending on the type of a business, being near competitors offers customers more choice that causes overwhelming and sales may not happen. So the different type centers should be at an admissible distance from each other.

%%%%%TODO
\noindent \textbf{Paper organization.} In \secref{defs}, we provide definitions and notations. In \secref{approx}, we present an $O(1)$ factor approximation
algorithm that guarantees $3\alpha /4$ separability. 
In \secref{constrained-dp} we present a
polynomial time algorithm for the constrained problem.
%We conclude in \secref{conclusions}.

\section{Problem and Definitions} \seclab{defs}
Let $\dist{\pnt,\query}$ denote the distance between points $\pnt, \query$ in the metric space $\Mspace$.
For a point $x \in \Mspace$ and a number $r \geq 0$ the ball $\ball{x,r}$ is
the set of points with distance at most $r$ from $x$, i.e.,
$\ball{x,r} = \{ \pnt \in \Mspace | \dist{x,\pnt} \leq r \}$ is the \emph{closed} ball of radius $r$ with center $x$.

In the $\alpha$-separated red-blue $(p+q)$-center clustering problem, we are given a set $\PntSet$ with $n$ points in $\Mspace$, integers $p > 0, q > 0$, and a real number $\alpha \geq 0$. For a given number $r \geq 0$, $p$ points $c_1, \ldots, c_p$ in $\Mspace$ (with possibly repeating points) called the red centers, and $q$ points $d_1, \ldots, d_q$ in $\Mspace$ (with possibly repeating points) called the blue centers, are said to be a \emphi{feasible solution} for the problem, with \emphi{radius of covering} $r$ if they satisfy,
\begin{itemize}
\item \textbf{Covering constraints:}
\[
\PntSet \subseteq \left( \bigcup_{i=1}^p \ball{c_i,r} \right) \bigcup \left( \bigcup_{j=1}^q \ball{d_j,r} \right).
\]

\item \textbf{Separation constraint:} For each $1 \leq i \leq p, 1 \leq j \leq q$, we have $\dist{c_i, d_j} \geq \alpha$, i.e., the red and blue centers are separated by at least a distance of $\alpha$.
\end{itemize}
The balls $\ball{c_i,r}, 1 \leq i \leq p$ are the \emphi{red balls} and $\ball{d_j,r}, 1 \leq j \leq q$ are the \emphi{blue balls}.
If there exists a feasible solution for a certain value of
$r$, such an $r$ is said to be feasible for the problem. The goal of the problem is to find the minimum possible value of $r$ that is feasible.

We denote this problem as the $(n,p \lor q, \alpha)$-problem. The $\lor$ in the
notation stresses the fact that union of the red balls and the blue balls cover $\PntSet$.
This is to be contrasted with the authors recent work  \cite{eskandari} where they consider the problem,
also considered previously by Kavand et. al. \cite{kavand} where \emph{both} the red and blue balls cover $\PntSet$. They denote that problem by the $(n, p \land q, \alpha)$ problem.

Let $r_{p \lor q,\alpha}(\PntSet)$ denote the optimal radius for this problem. When $\PntSet, p, q, \alpha$ are clear from context we will also denote this by $\ropt$ for brevity. Let $r_{k}(\PntSet)$ denote the optimal $k$-center clustering radius, for all $k \geq 1$. Notice that
the centers in the $k$-center clustering problem can be
any points in $\Mspace$, not necessarily belonging to
$\PntSet$. If that is the requirement, the problem is 
the \emph{discrete} $k$-center clustering problem.

We will always be concerned with 
$\Mspace = \Re^d$. We let $\PntSet = \{\pnt_1, \ldots, \pnt_n\}$ where,
$\pnt_i = (\pnt_{i1}, \pnt_{i2}, \ldots, \pnt_{id})$. We also consider the \emph{constrained} $\alpha$-separated red-blue $(p+q)$-center clustering problem (when $\Mspace = \Re^d$). Here, we are given
a line $\ell$ and all the centers are constrained to lie on $\ell$. Without loss of generality, we assume that $\ell$ is the $x$-axis. This can be achieved by an appropriate transformation of space. Moreover, we will
use the same notation for the optimal radii and centers. For the constrained
problem we need some additional definitions and notations.  
%We assume that no two points in $\PntSet$ have the same distance from $\ell$. (This general position assumption can however be removed.)
For each point $\pnt_i \in \PntSet$, we consider the set of points on the line ($x$-axis) such that the
ball of radius $r$ centered at one of those points can cover $\pnt_i$. This
is the intersection of $\ball{\pnt_i, r}$ with the $x$-axis.%, see \figref{circ}.
 Assuming this intersection is not empty, let the interval be 
$\Int_i(r) = [\lept_i(r), \rept_i(r)]$. Denote the set
of all intervals as $\IntSet(r) = \{ \Int_1(r), \ldots, \Int_n(r) \}$ where we 
assume that the numbering is in the sorted order of intervals: those
with earlier left endpoints are before, and for the same left endpoints
the one with earlier right endpoint occurs earlier in the order. Notice that feasibility of radius $r$ means that there
exists a hitting set for the set of intervals $\IntSet(r)$, that can be partitioned into
the red centers and the blue centers satisfying the separation constraint.
%\setlength{\intextsep}{-3pt}%
%\begin{wrapfigure}{r}{0.5\textwidth}
%  \centering
%  \captionsetup{justification=centering}
%    \includegraphics[width=0.50\textwidth]{circle}
%    \caption{The functions $\lept_i(r), \rept_i(r)$}
%    \figlab{circ}%
%\end{wrapfigure}
The interval
endpoints $\lept_i(r), \rept_i(r)$ can be computed by solving the equation,
$(x-\pnt_{i1})^2 + \pnt_{i2}^2 + \ldots + \pnt_{id}^2 = r^2,$ for $x$. Thus, they
are given by $\lept_i(r) = p_{i1} - \sqrt{r^2 - \sum_{j=2}^d p_{ij}^2}$,
and $\rept_i(r) = p_{i1} + \sqrt{r^2 - \sum_{j=2}^d p_{ij}^2}$. It is easy to see
that for the range of $r$ where the intersection is non-empty, $\lept_i(r)$ is 
a strictly decreasing function of $r$ and $\rept_i(r)$ is a strictly increasing function
of $r$.

\noindent \textbf{Model of computation.} We remark that our model of computation is the Real RAM model, where the usual arithmetic operations are assumed to take $O(1)$ time.

\section{Approximation algorithms}
\seclab{approx}%
Our goal here is to show that there is a 
polynomial time algorithm that computes a
bi-criteria approximation to the $(n, p \lor q, \alpha)$ problem for a given point set
$\PntSet$. The algorithm will find $p$ red
and $q$ blue centers such that:
(i) The distance between each red and each blue
center is at least $3\alpha/4$, and, (ii) the covering radius will be at most $8 r^*$, where
$r^*$ is the optimal covering radius. The algorithm works differently when $r^* < \alpha/8$ (\secref{largealpha}) than when $r^* \geq \alpha/8$ (\secref{smallalpha}). Notice that this is not known to us at the beginning, so we explain how to combine the two in \secref{finalalgo}. We first discuss the case when $r^*$ is large(i.e., $r^* \geq \alpha/8$), and then
proceed to the more complex case when $r^*$ is small ($r^* < \alpha/8$). An easy observation is that
$r^* = r_{p \lor q, \alpha}(\PntSet) \geq r_{p+q}(\PntSet)$. To see this, notice that in both the $(p+q)$-center problem and the $(n, p \lor q, \alpha)$ problem we cover $\PntSet$ by balls centered at $p + q$
points. However the $(n, p \lor q, \alpha)$ problem has additional constraints, as such the covering radius can only be larger.

\subsection{The case when \texorpdfstring{$r^* \geq \alpha/8$}{rga}}
\seclab{smallalpha}%
The algorithm first computes a $2$-approximation to the $(p+q)$-center clustering
problem. We briefly review this. This is via the same idea as Gonzalez's algorithm \cite{gonzalez}. We start with any point of $\PntSet$ as the first center. Then, 
until $(p+q)$ points have been found, we select
the furthest center in $\PntSet$ from the current ones. 
While the analysis of Gonzalez \cite{gonzalez} was
for the discrete $k$-center problem, it works
without any changes for the $k$-center problem
where the centers are not restricted to be
points of $\PntSet$, and we get a $2$-approximation
to $r_k(\PntSet)$.

At the end of this step we have found points
$x_1, x_2, \ldots, x_{p+q} \in \PntSet$ such that the radius of covering of $\PntSet$ using these points is at most $2 r_{p+q}(\PntSet)$. Then, we select a maximal subset of these points that are at least $3\alpha/4$ away from each other. Suppose that, possibly with some renaming, these points are $x_1, x_2, \ldots, x_t$ where $1 \leq t \leq (p+q)$.
We place $(p+q)$ red and blue centers among these points. In doing so it does not matter which points are which color, except that there are at most $p$ red points and $q$ blue points. If there is at least 1 red and 1 blue point, then we can increase the number of red (resp. blue) points to $p$ (resp. $q$) by possibly co-locating some points. However, if $t=1$, assume it is red; then, one can choose all the $p$ red points at $x_1$, and choose all the $q$ blue points on the surface of the ball $\ball{x_1, 3\alpha/4}$. The separation claim that
red and blue centers are spaced at least $3\alpha/4$ apart is clear by construction. The following lemma proves the claim about the covering radius.
\begin{lemma}
The covering radius using the centers constructed is at most $8r^*$.
\end{lemma}

\begin{proof}
By the guarantee of Gonzalez's algorithm we 
have that,
$\PntSet \subseteq \bigcup_{i=1}^{p+q} \ball{x_i, 2r_{p+q}(\PntSet)}$. As such, for the points $x_1, \ldots, x_t$ we can say
$\PntSet \subseteq \bigcup_{i=1}^t \ball{x_i, 2r_{p+q}(\PntSet) + 3\alpha/4}$, by the triangle inequality.
(Adding more centers for the $t=1$ case can only decrease the covering radius.)
The covering radius is thus bounded by,  $2r_{p+q}(\PntSet) + 3 \alpha/4 \leq 2 r^* + 6 r^* = 8r^*,$ where the first inequality follows from our observation above, and the second from the
assumption that $r^* \geq \alpha/8$. The claim is proved.
\end{proof}
Notice that the algorithm itself makes no use
of the knowledge of $r^*$. The claim about the covering radius is however only valid when $r^* \geq \alpha/8$. Notice that it can be determined by clustering each point to its nearest center. We summarize the result in the following lemma,
\begin{lemma}
\lemlab{algosmallalpha}%
There is a polynomial time algorithm that  returns a set of $p$ red centers $c_1, \ldots, c_p$, $q$ blue centers $d_1, \ldots, d_q$, and a number $r$, such that,
\[
\PntSet \subseteq \left( \bigcup_{i=1}^p \ball{c_i,r} \right) \bigcup \left( \bigcup_{j=1}^q \ball{d_j,r} \right),
\]
and $\dist{c_i,d_j} \geq 3\alpha/4$ for $1\leq i\leq p, 1 \leq j \leq q$. If $r^* \geq \alpha/8$, then $r \leq 8 r^*$.
\end{lemma}

\subsection{The case when \texorpdfstring{$r^* < \alpha/8$}{rla}}
\seclab{largealpha}%
We first present a decision procedure that takes a number $R \geq 0$ as input and returns $\true/\false$. If it returns $\true$ it also finds a feasible solution with a bi-criteria guarantee: covering radius of at most $2R$ and separability of at least $3\alpha/4$. Then, we show that when $r^* < \alpha/8$, our decision procedure will definitely return $\true$ for some $R \leq 2r^*$.

\noindent \textbf{The decision procedure.} 
We first 
construct a neighborhood graph $G$ on $\PntSet$ as vertex set, by connecting two points $p, q \in \PntSet, p \neq q$ by an edge if their distance is at 
most $3 \alpha/4$. Then, we compute the connected components of $G$. Let the connected components be $C_1, \ldots, C_m$. Now, for each connected component $C_i$, we run the following scooping algorithm. Start with an arbitrary point $x_{i_1} \in C_i$.
Suppose that points $x_{i_1}, \ldots, x_{i_j}$
have already been selected in $C_i$. If $C_i \subseteq \bigcup_{k=1}^j \ball{x_{i_k},2R}$
we stop, else we select the next point $x_{i_{j+1}}$ as any point in $C_i \setminus \left(\bigcup_{k=1}^j \ball{x_{i_k},2R}\right).$ Let $n_i$ be the number of centers thus found. We now have positive integers $n_1, \ldots, n_m$ and we want to solve the following problem: Does there exist a partition $A \cup B = \{ 1, \ldots, m \}$ such that, $\sum_{j \in A} n_j \leq p$ and $\sum_{j \in B} n_j \leq q$? Notice that this problem is NP-hard when the input is the set of integers $n_1, \ldots, n_m, p, q$. However, since our input size is at least $n$, this problem can be solved in polynomial time by a dynamic programming algorithm specified by the following recursion. Let $T[k,a,b]$ represent a table of $\true / \false$, where
$0 \leq k \leq m, 0 \leq a \leq p, 0 \leq b \leq q$ and the table entry is $\true$ iff the numbers $n_1, \ldots, n_k$ can be partitioned into two parts one of which sums to at most $a$ and the rest to at most $b$. The recursive definition is as follows:
\[
T[k,a,b] = 
\begin{cases}
\false \text{ if } \sum_{1 \leq i \leq k} n_i > a + b \\ \true \text{ if } k=0 \\
B_1 \lor B_2 \text{ otherwise},
\end{cases}
\]
where the Boolean $B_1$ is defined as,
\[
B_1 \gets (n_k \leq a) \land T[k-1,a-n_k,b],
\]
and similarly $B_2$ is defined as,
\[
B_2 \gets (n_k \leq b) \land T[k-1,a,b-n_k].
\]

If the above partitioning problem is $\true$, i.e., $T[n,p,q]$ is $\true$, the algorithm returns $\true$, else it returns $\false$. The partitioning can also be computed by the dynamic programming. Thus, when it returns $\true$ the algorithm also finds a partitioning of the connected components into 2 parts, one, whose associated integers $n_i$ add up to at most $p$, and the other, whose associated integers add up to at most $q$. We color the centers associated with the components in the former class, red, and those with the other, blue. Then, clearly we have output at most $p$ red and at most $q$ blue centers. By possibly co-locating some centers, we can return exactly $p$ red and $q$ blue centers. We can now show the following lemma,
\begin{lemma}
\lemlab{returntype}%
If the procedure above returns $\true$, it finds $p$ red and $q$ blue centers such that the distance between each red and each blue center is larger than $3\alpha/4$, and they cover $\PntSet$ using a radius of at most $2R$. 
\end{lemma}
\begin{proof}
If the procedure returns $\true$, clearly it found a partition of the components $C_1, \ldots, C_m$ into 2 parts, such that the components in the one part marked with red centers can be covered by balls of radius $2R$ around the at most $p$ such centers.
Similarly the blue centers cover all the components with the blue centers, with balls of radius $2R$. Each point of $\PntSet$ is in one of these components, so all of $\PntSet$ is covered. For the separation guarantee, notice that our algorithm returns centers that are points of the components themselves. But given any two points across components, they are clearly separated by a distance of more than $3\alpha/4$ otherwise those points would be in the same component of $G$. So,
each red center and each blue center (or even two similarly colored centers in different components) are separated by more than $3 \alpha/4$.
\end{proof}

Our next lemma, and its proof, are actually the crux of the algorithm. First, a definition. We say a number $D \geq 0$ is an \emphi{interpoint distance} if there exist points $\pnt_i, \pnt_j \in \PntSet$ such that $D = \dist{\pnt_i, \pnt_j}$. Notice that $0$ is an interpoint distance.

\begin{lemma}
\lemlab{smallropt}%
If $r^* < \alpha/8$, then there exists an interpoint distance $D$ such that, $D \leq 2r^*$ and if $D \leq R$ the algorithm returns $\true$.
\end{lemma}
\begin{proof}
Assume that $r^* < \alpha/8$. Consider an optimal solution: $p$ red centers $c_1, \ldots, c_p$ and $q$ blue centers $d_1, \ldots, d_q$ with covering radius $r^*$. For a point $x \in \PntSet$ if $x$ is covered by a red ball we say it is red. If it is covered by a blue ball we say it is blue. (A point can potentially have both the colors.) We claim that, if $x, y \in \PntSet$ are any two points, then if $\dist{x,y} \leq 3\alpha/4$, then $x,y$ cannot be of different colors.
To see this we proceed by contradiction. Suppose, wlog, $x$ is red and $y$ is blue. Let $c_i$ be a red center covering $x$ and $d_j$ a blue center covering $y$. By the separation constraint, $\dist{c_i, d_j} \geq \alpha$. On the other hand, 
$\dist{c_i, x} \leq r^* < \alpha/8$ and similarly, $\dist{d_j, y} < \alpha/8$. Then, by the triangle inequality,
\[
\alpha \leq \dist{c_i, d_j} \leq \dist{c_i, x} + \dist{x,y} + \dist{y, d_j} < \alpha/4 + \dist{x,y},
\]
and this leads to $\dist{x,y} > 3\alpha/4$, a contradiction. For $x = y$ the above implies that each point is a unique color. Next we claim that, for each component $C_i$, either all points are red or all are blue.
To see this, suppose that the claim is false. Then, there exist points $x, y \in C_i$ such that $x$ is red and $y$ is blue. Since $C_i$ is connected, there is a path from $x$ to $y$ in $C_i$ and since the endpoints of this path have different color, there must be an edge in this path such that its endpoints $x', y'$ have different color, say $x'$ is red and $y'$ is blue. But, this contradicts the last claim, since $\dist{x',y'} \leq 3\alpha/4$.
The next claim is easy to see, since its proof closely follows the proof of our first claim above:
No ball covers points in two different components. Due to the claims above, we may assume that we can partition the set of $p+q$ centers into $m$ parts, each part is the set of the centers covering a specific component $C_i$. Let the number of centers whose balls cover $C_i$ be $s_i$, for $i=1,\ldots, m$. Then, the sum of these numbers that correspond to the red centers sum up to $p$, and those for the blue centers sum up to $q$.  Now consider the $p+q$ balls of radius $r^*$ covering $\PntSet$. For each such ball $B$, it covers a set of points $X \subseteq \PntSet$. Consider two points $p, q \in X$ among these points that are a furthest pair. Clearly $X$ is also covered by the ball $\ball{x,\dist{x,y}}$. Out of the $p+q$ balls we consider the ball that gives us the furthest such pair, i.e., the ball that gives us the furthest apart diametral pair. Call this furthest interpoint distance $D$. We claim that $D \leq 2r^*$ and that if $D \leq R$, the algorithm must return $\true$. The first is easy to see since the diametral pair is inside one of the balls $B$ of radius $r^*$. To see the second statement, consider a component $C_i$ and one of the balls $B$ out of the $n_i$ balls that cover $C_i$. Clearly, the ball of radius $D$ centered at any point of $\PntSet \cap B$ also covers $\PntSet \cap B$. Then, assuming $D \leq R$, if we follow the scooping procedure of our decision algorithm, by Gonzalez's guarantee \cite{gonzalez}, the number of balls generated $n_i$ is at most $s_i$.
As such there will be a partition of $\{1,\ldots,m\}$ such that the $n_i$ for the indexes corresponding to red centers will sum up to at most $p$ and those in the others to at most $q$. In other words our partitioning algorithm will return $\true$ and so will the decision algorithm.
\end{proof}

\noindent \textbf{The algorithm for $r^* < \alpha/8$.} The algorithm is simple. We enumerate and try all the $O(n^2)$ interpoint distances. Either the decision algorithm returns $\false$ for all of them, or we return the set of centers returned by the smallest interpoint distance $D$ for which it returns $\true$. We conclude the following result.

\begin{lemma}
\lemlab{algolargealpha}%
There is a polynomial time algorithm that either terminates without returning anything, or it returns a set of $p$ red centers $c_1, \ldots, c_p$, $q$ blue centers $d_1, \ldots, d_q$ all in $\PntSet$, and a number $r$, such that,
\[
\PntSet \subseteq \left( \bigcup_{i=1}^p \ball{c_i,r} \right) \bigcup \left( \bigcup_{j=1}^q \ball{d_j,r} \right),
\]
and $\dist{c_i,d_j} > 3\alpha/4$ for $1\leq i\leq p, 1 \leq j \leq q$. If $r^* < \alpha/8$, then the algorithm does return the centers mentioned, and moreover, $r \leq 4 r^*$.
\end{lemma}

\begin{proof}
The claim about the ``return type'' of the algorithm follows from \lemref{returntype} and the structure of the algorithm. If $r^* < \alpha/8$, then by \lemref{smallropt}, there exists an interpoint distance $D$ such that $D \leq 2r^*$ and for $D \leq R$ the decision algorithm returns true along with the centers satisfying the claimed separation guarantee. Since we check all interpoint distances as values for $R$, and return the centers for the smallest one that returns $\true$, the given claim is true. Clearly the covering balls have radius at most $2D \leq 4r^*$ in this case.
\end{proof}

\subsection{Putting things together}
\seclab{finalalgo}%
This section just consolidates the results of \secref{smallalpha} and \secref{largealpha}. The algorithm runs the algorithms of both the sections, i.e., the ones detailed in \lemref{algosmallalpha}
and \lemref{algolargealpha}. Either both return centers or only one of them (\lemref{algosmallalpha})
does. Notice that in both the cases, to determine the covering radius, we can just do the usual nearest-neighbor clustering, i.e., each point is clustered to its closest center. We choose the one with the best covering radius. Since, one works for $r^* \geq \alpha/8$ and the other in the complementary case $r^* < \alpha/8$, and we choose the one with the smallest clustering radius, we get at least the stricter of the guarantees. We get the following result.
\begin{theorem}
\thmlab{final-approx}%
There is a polynomial time algorithm that given an instance of the $(n, p \lor q, \alpha)$ problem over a set of points $\PntSet$ with $\cardin{\PntSet} = n$, returns a set of $p$ red centers $c_1, \ldots, c_p$, $q$ blue centers $d_1, \ldots, d_q$, and a number $r$ such that,
\[
\PntSet \subseteq \left( \bigcup_{i=1}^p \ball{c_i,r} \right) \bigcup \left( \bigcup_{j=1}^q \ball{d_j,r} \right),
\]
and $\dist{c_i,d_j} \geq 3\alpha/4$ for $1\leq i\leq p, 1 \leq j \leq q$. Moreover, $r \leq 8 r^*$.
\end{theorem}

\section{The constrained problem}

\seclab{constrained-dp}
In this section, we consider the constrained $(n, p \lor q, \alpha)$  problem in which the centers are restricted to lie on a given line $\Line$. Without loss of generality, we assume that $\Line$ is the $x$-axis.
First, we consider the decision problem and present a dynamic programming algorithm to decide if the constrained $(n, p \lor q, \alpha)$  problem has a feasible solution with a given covering radius $r$. Next, we will find a discrete set of candidates for possible values of $r^*$. By performing a binary search over this set of candidates and check their feasibility via the dynamic programming algorithm, we can find the optimal solution. In this section, for those points that are on the $x$-axis, be a slight abuse of notation, we may use a point instead of its $x$-coordinate.
\subsection{Decision Problem}
For a given radius $r$, we want to decide if the constrained $(n, p \lor q, \alpha)$  problem has a feasible solution with covering radius $r$.
We find a finite set of candidates for possible locations of centers on $\Line$ and use it for designing a dynamic programming algorithm to find a feasible solution, if there exists one.

Consider the endpoints of the intervals in $\IntSet$, i.e., $\lept_i(r)$ and $\rept_i(r)$, for $1\leq i\leq n$. Let $x_1(r), x_2(r), \dots ,x_{2n}(r)$ be the sorted order of these endpoints on $\Line$ and $\CandSet(r)=\{x_i(r) , x_i(r)+\alpha | 1\leq i\leq 2n, x_1 (r)\leq x_i(r)+\alpha \leq x_{2n}(r)\}$. We show that if $r$ is feasible, there exists a feasible solution which is a subset of $\CandSet(r)$. Thus, we can search over $\CandSet(r)$ for a solution.

First, we need a definition.
Let $U=\{u_1, u_2, \dots, u_{p+q}\}$ be a feasible solution with covering radius $r$. It is called \emphi{standard} if it has all the four following properties:
\begin{inparaenum}
\item $U\subset \CandSet(r)$.
\item $u_1$ and $u_{p+q}$ are on the endpoints.
\item Any two consecutive same color centers are on the endpoints.
\item If $u_i$, $2\leq i\leq p+q-1$, is not on an endpoint, its adjacent centers ($u_{i-1}$ and $u_{i+1}$) are on the endpoints and $u_i-u_{i-1}=\alpha$.
\end{inparaenum}
We need only the first property of a standard solution for our dynamic programming algorithm.
 The other properties will be used to find a set of finite number of candidates for the radius of covering in the next section. A feasible solution can be converted to standard form by the following approach:

\noindent {\bf Converting a given solution to standard form:}
Let $U=\{u_1, u_2, \dots , u_{p+q}\}$ (which are sorted from left to right on $\Line$) be a feasible solution with radius of covering $r$. If $U\subset \{x_1(r), x_2(r), \dots , x_{2n}(r)\}$, we are done. Now assume that there is at least one center in $U$ that is not on an endpoint.
%%NK - discussion on faces
The feasibility problem is equivalent to finding a hitting set for the intervals in $\IntSet(r)$ that can be divided into two subsets satisfying the separation constraints. As is standard in such hitting set problems, we look at the \emph{faces} of the arrangement of the 
intervals. Some of these faces might be open intervals, or half open intervals, or even singleton points. All faces are disjoint by definition. If $F$ is such a face, then a point in the closure $\cl{F}$ of $F$ will hit at least the same intervals as points in $F$ hit. To compute all the face closures in the arrangement of the $\IntSet(r)$, we retain all the
consecutive intervals $[x_i(r), x_{i+1}(r)]$ that do not lie outside any of the intervals $\Int_i(r)$. This can be done by a simple line sweep algorithm. Notice that there are only $O(n)$ such face closures. Also, we may assume each such face closure contains at most one center since otherwise it is easy to see that they can be ``fused'' together and colored in a way still preserving the separation constraints with the other centers.
%%%
The face closure that contains $u_i$ is denoted by $[x_{i,1}(r),x_{i,2}(r)]$, where $x_{i,1}(r)$ and $x_{i,2}(r)$ are the endpoints of some intervals (i.e., $a_j(r)$ or $b_j(r)$). To convert $U$ to standard form, we construct a set of $p+q$ points belonging to $\CandSet(r)$ that is a feasible solution with covering radius $r$ with the desired properties. It will be done in two phases: in the first phase, we shift as many canters as we can, to the endpoints of the intervals (i.e., $\lept_j(r), \rept_j(r)$). Clearly only some of the centers are allowed to shift, because of covering and separation constraints. In the second phase, the centers that were not shifted to the endpoints in the first phase, will be shifted to some points of $\CandSet (r)$ such that the solution remains feasible. Moreover, in the second phase, we will construct a solution in which, if a center is not on an endpoint, its adjacent centers are on the endpoints.

{\bf Phase 1:} First of all, if $u_1$ (resp. $u_{p+q}$) does not lie on an endpoint,  let $u_1=x_{1,1}(r)$ (resp. $u_{p+q}=x_{p+q,2}(r))$.
For each $u_i$, $2\leq i\leq p+q-1$, that does not lie on an endpoint, if $u_i$ and $u_{i-1}$ are the same color, let $u_i=x_{i,1}(r)$.
Otherwise, consider $u_{i+1}$. If $u_i$ and $u_{i+1}$ are the same color, let $u_i=x_{i,2}(r)$. Since each face closure contains at most one center we never cross other centers while moving centers.
\\
After Phase 1, a sequence of consecutive centers with the same color lies on the endpoints.

%{\bf Phase 2:} For all $i, \; 2\leq i\leq p+q-1$, that $u_i$ is the leftmost center which is not on an endpoint after Phase 1, do the 
{\bf Phase 2:} Let $u_i, 2 \leq i \leq p +q -1$ be the leftmost center that is not on an endpoint after Phase 1, if there exists at least one such center, else we are done with this Phase. Do the
following process for $u_i$. First note that $u_{i-1}$ is on an endpoint and $u_i$ should have different color with both its neighbours, $u_{i-1}$ and $u_{i+1}$ (otherwise, it had been shifted to an endpoint in Phase 1). So $u_{i-1}$ and $u_{i+1}$ have the same colors.
 We shift $u_i$ to the left until it hits either $x_{i,1}(r)$ or $u_{i-1}+\alpha$. In the other words, if $x_{i,1}(r)-u_{i-1}\geq \alpha$, let $u_{i}=x_{i,1}(r)$, otherwise, $u_{i}=u_{i-1}+\alpha$.
Next, for satisfying property 4, if $u_{i}$ is not on an endpoint yet, we need to consider $u_{i+1}$ to move it on an endpoint.

First, if $u_{i+1}$ is already on an endpoint (including the case $i+1=p+q$), we are done with $u_i$. Otherwise, $i+1<p+q$ and centers $u_{i+1}$ and $u_{i+2}$ are not the same color (due to Phase 1).
Consider the four centers $u_{i-1}$, $u_i$, $u_{i+1}$, and  $u_{i+2}$. Clearly, $u_{i-1}$ and $u_{i+1}$ are the same color and $u_{i}$ and $u_{i+2}$ are the same color. We invert the colors of $u_i$ and $u_{i+1}$. This does not change the number of red and blue balls. Moreover, the solution remains feasible after inverting the colors. In this new order of centers,  $u_{i-1}$ and $u_i$ are the same color and $u_{i+1}$ and $u_{i+2}$ are the other color. Next,
let $u_i=x_{i,1}(r)$ and $u_{i+1}=x_{i+1,2}(r)$.
After this displacement, the new solution satisfies the separation constraint, since the distance between $u_i$ and $u_{i+1}$ is greater than their previous distance, which was also at least $\alpha$ because of feasibility of $U$. Now, $u_{i+1}$ is also on an endpoint. 
Repeat Phase 2 again until done. %Then we repeat this phase for $i=i+2$, if $i+2\leq p+q-1$.

In the new feasible solution obtained from the above approach, if a center, say $u_i$, does not lie on an endpoint, it should have a color different from its two neighbors, $u_{i-1}$ and $u_{i+1}$, that therefore have the same color. In addition, in the sequence of three consecutive red-blue centers, $u_{i-1}$, $u_i$, and $u_{i+1}$, $u_{i-1}$ and $u_{i+1}$ must be on endpoints and $u_i$ is at a distance of $\alpha$ from $u_{i-1}$.
 Thus, all centers belong to $\CandSet(r)$. Thus, we have proved the following lemma:

\begin{lemma}
If $r$ is a feasible radius for the constrained $(n, p \lor q, \alpha)$  problem, there exists a feasible solution with covering radius $r$ such that the centers belong to $\CandSet(r)$.
\lemlab{candset}
\end{lemma}

%Now we have a finite set of possible locations for the centers. By having this finite set, we present a dynamic programming algorithm to answer the decision problem.
Now, we present the dynamic programming algorithm for the decision problem.

For a given $r$, we sort $\CandSet(r)$ from left to right to obtain $\cpt_1, \cpt_2,\dots , \cpt_m$, where $m\leq 4n-1$. The index of the first point after $\cpt_j$ in $\CandSet(r)$ with a distance of at least $\alpha$ from $\cpt_j$, is denoted by $NEXT(j)$. If there is no such point, let $NEXT(j)=m+1$. Note that the intervals $\Int_i(r)$ are sorted in non-decreasing order of their left endpoints, $\lept_i(r)$, which leads to an order on points of $\PntSet$. Let $\pnt_1, \ldots, \pnt_n$ be the points ordered by inheritance from the ordering of $\Int_i(r)$. For $0\leq i\leq n$, $0\leq p'\leq p$, $0\leq q'\leq q$ and $1\leq j\leq m+1$, let $TR$ (resp. $TB$) is a 4-dimension $\true$/$\false$ array such that $TR[i,p',q',j]=\true$ (resp.  $TB[i,p',q',j]=\true$), if we can cover the last $i$ points, $p_{n-i+1}, p_{n-i+2},\dots , p_n$, with $p'$ red balls and $q'$ blue balls when the leftmost center is a red (resp. blue) center on $\cpt_j$, and such that the distance between blue centers and red centers is at least $\alpha$. Otherwise $TR[i,p',q',j]=\false$ (resp.  $TB[i,p',q',j]=\false$).

Given the definition of tables $TR$ and $TB$ for a given radius $r$, the value of the following Boolean expression determines if there is a solution or not:

$$\left(\bigvee_{j=1}^m TB[n,p,q,j]\right )\bigvee \left ( \bigvee_{j=1}^m TR[n,p,q,j]\right )$$

This expression returns true if  for some $j$, either $TR[n,p,q,j]$ or $TB[n,p,q,j]$ is true.  We know that $TR[n,p,q,j]=\true$ (resp. $TB[n,p,q,j]=\true$), if we can cover all points $p_1, p_{2},\dots , p_n$ with $p$ red balls and $q$ blue balls when the leftmost center is red (resp. blue) located at $\cpt_j$ such that the separation constraints hold. In other words, we should hit all the intervals in $\IntSet$ by at least one red or blue center by trying all possible starting locations and color for the leftmost center.

We use a recursive approach to compute the tables entries. First, we find the initial values.
If all intervals have been already hit, i.e., $i=0$, we should return $\true$. So for all  $1\leq j\leq m+1$, $0\leq p'\leq p$ and $0\leq q'\leq q$,  we have: $TR[0,p',q',j]= TB[0,p',q',j]=\true$.
If there are not any red or blue centers
to put while some unhit intervals remain, we return $\false$. Thus, for all  $1\leq j\leq m$, $i>0$,  $TR[i,0,0,j]= TB[i,0,0,j]=\false$.
If we have already passed over all
centers but any unhit intervals remain,
we return $\false$. So for all  $i>0$ and $p', q' \geq 0$,  $TR[i,p',q',m+1]= TB[i,p',q',m+1]=\false$.
If there is no more red (resp. blue) center, we will no longer be able to place a red (resp. blue) center at $\cpt_j$, so for all  $1\leq j\leq m+1$, $i>0$ and $p', q' \geq 0$,  $TR[i,0,q',j]= TB[i,p',0,j]=\false$.

For a given $p'$ and $q'$, $TR[i,p',q',j]$ (resp. $TB[i,p',q',j]$) can be supposed as the $(i,j)$-th entry of a $(n+1)\times (m+1)$ matrix. We start with initializing the entries of $TR[i,0,0,j]$,  $TB[i,0,0,j]$, $TR[i,0,1,j]$,  and  $TB[i,1,0,j]$.  Matrices $TR[i,1,0,j]$ and $TB[i,1,0,j]$ can be easily computed by placing a center at $\cpt_j$ and deciding if $\ball{\cpt_j,r}$ can cover all points $\pnt_{n-i+1}, \pnt_{n-i+2},\dots , \pnt_n$, because we have just one center.

 Now for each pair $(p',q')$ (starting from $(1,1)$), suppose that we have already computed all entries of these matrices: $TR[i,p'-1,q',j]$ and $TB[i,p'-1,q',j]$ (resp. $TR[i,p',q'-1,j]$) and $TB[i,p',q'-1,j]$).
%%%%%%%%%%%%%%%%%
To compute $TR[i,p',q',j]$ (resp.  $TB[i,p',q',j]$), first of all, let $p_s, p_{s+1},\dots , p_{s'}$ be the points that are covered by the ball $\ball{\cpt_j,r}$. If $s>n-i+1$, $TR[i,p',q',j]=TB[i,p',q',j]=\false$. This is so, because if points $p_{n-i+1},\dots , p_{s-1}$ cannot be covered by $\ball{\cpt_j,r}$, then none of the other balls would be able to cover those points since for all $k>j$, $\cpt_k>\cpt_j$ and if $\ball{\cpt_j,r}$ cannot cover those points, neither can $\ball{\cpt_k,r}$.
If $s\leq n-i+1$, $TR[i,p',q',j]$ and $TB[i,p',q',j]$ can be computed by the recursive formulae:
\begin{align*}
&TR[i,p',q',j]=\left (\bigvee_{k=j+1}^m  TR[i',p'-1,q',k] \right ) \bigvee \left ( \bigvee_{k=NEXT(j)}^m TB[i',p'-1,q',k]\right ) \\
&TB[i,p',q',j]=\left (\bigvee_{k=j+1}^m  TB[i',p',q'-1,k]\right ) \bigvee\left ( \bigvee_{k=NEXT(j)}^m TR[i',p',q'-1,k]\right )
\end{align*}
where $i'=\left | \{p_{n-i+1}, p_{n-i+2}, \dots , p_n \} \setminus \{p_s, p_{s+1},\dots , p_{s'}\}\right |$
\footnote { Since $s\leq n-i+1$,  by removing the points $p_s, p_{s+1},\dots , p_{s'}$ from the set of uncovered points,  $\{p_{n-i+1}, p_{n-i+2}, \dots , p_n \}$, we will obtain $i'$ consecutive points  $\{p_{n-i'+1}, p_{n-i'+2}, \dots , p_n \}$.}.

As per the definition
of the table $TR[i,p',q',j]$ (resp. $TB[i,p',q',j]$), the next center is red (resp. blue)
 at $\cpt_j$ and may cover some uncovered points. We remove those
points from the remaining points (i.e., $\{p_{n-i+1}, p_{n-i+2}, \dots , p_n \}$). Then we try all possible starting locations and color for the first unplaced center. Finally, all entries of $TR[i,p,q,j]$ and $TB[i,p,q,j]$ can be computed and we can decide if the problem has a solution or not.

{\bf Analysis:}
Note that $|\CandSet(r)|=O(n)$  and computing the candidate values for centers
can be done in $O(n)$ time. Moreover, the successor
points $NEXT(j)$ can be computed in total
$O(n \log n)$ time by first sorting $\CandSet(r)$
and searching for $\cpt_j+\alpha$ in the sorted list.
There are in total $O(n^2 p q)$ entries
to be filled since $m=|\CandSet(r)|=O(n)$.
The points that remain uncovered after putting a center at $\cpt_j$, can be easily found in $O(n)$ time.
Moreover, each entry has $O(n)$ Boolean terms and can be looked up in $O(n)$ time. Overall, we will
take $O(n^3 p q)$ time. Therefore, we have the following theorem:

\begin{theorem}
\thmlab{feasibility-main}%
For the constrained $(n, p \lor q, \alpha)$ problem, it can be decided if a given $r$ is feasible in time
$T_{DP}(n,p,q)=O(n^3 p q)$. Moreover, if $r$ is feasible,
a feasible solution with covering radius $r$ can also be
computed in the same time.
\end{theorem}

%Note that for computing a feasible solution, we need to find a value of $j$ for which  $TR[n,p,q,j]=\true$ or $TB[n,p,q,j]=\true$, say $TR[n,p,q,j]=\true$. So we put a red center at $\cpt_j$. Next, we choose one of the clauses in the expression $TR[n,p,q,j]$ with the truth value, say $TB[i,p',q',k]=\true$. We put a blue center at $\cpt_k$. We continue following the clauses with truth value to reach $i=0$. This can be done in $O(n(p+q))$ time.

\subsection{Candidate values for \texorpdfstring{$r$}{r}}
For finding a discrete set of candidates for the optimal radii, we will prove a property of the optimal solution.
For this purpose, we need  some notations and definitions.

Let $x_s(r)$ be any endpoint of the interval $\Int_j(r)$ and $x_t(r)$ be an endpoint of $\Int_k(r)$. The distance between $x_s(r)$ and $x_t(r)$ is called \emphi{exceptional}, if it satisfies the following properties:
\begin{inparaenum}
\item $\pnt_j$ and $\pnt_{k}$ are equidistant from $\Line$.
\item $\pnt_{j1}-\pnt_{k1}=\alpha \; \text{or} \; 2\alpha$.
\item Either $x_s(r)=\lept_j(r), x_t(r)=\lept_k(r)$ or $x_s(r)=\rept_j(r), x_t(r)=\rept_k(r)$.
\end{inparaenum}
Otherwise, the distance is \emphi{non exceptional}.

%Moreover, an exceptional distance is called $\alpha$-\emphi{exceptional}, if $\pnt_{j1}-\pnt_{k1}=\alpha $ otherwise it is called $\alpha$-\emphi{exceptional}.

If $\pnt_j$ and $\pnt_{k}$ are equidistant from $\Line$, we have $\lept_{j}(r)-\lept_{k}(r)=\rept_j(r)-\rept_{k}(r)=\pnt_{j1}-\pnt_{k1}$ for all values of $r$. 
Thus exceptional distances do not change when the radius $r$ changes.

\begin{lemma}
 Let $r$ be a feasible radius for the constrained $(n,p \lor q,\alpha)$ problem. If any non exceptional distance between two endpoints is not
 $0$, $ \alpha$ or $2\alpha $,  then the constrained $(n,p\lor q,\alpha)$ problem has a feasible solution with radius less than  $r$.
 \lemlab{candidate}
\end{lemma}
\begin{proof}
We show that there is a real number $0<\epsilon<r$ such that the constrained $(n,p\lor q,\alpha)$ problem has a feasible solution with radius of covering  $r-\epsilon$. To this end, we obtain a set of centers, $\bar{U}$, from the given standard solution ($U$) and show that the set of balls centered at the points in  $\bar{U}$ with  radius $r-\epsilon$ is a feasible solution for the problem. First, we need to modify $U$ to find a feasible solution with the property that any two non exceptional consecutive blue and red centers are at a distance  strictly greater than $\alpha$\remove{NK: This is confusing: (not exactly $\alpha$)}. Then we use it for finding a solution, $\bar{U}$, with radius of covering $r-\epsilon$ ($\epsilon$ will be fixed later).

Let $U$ be a standard solution with radius of covering $r$.
For any two consecutive red and blue centers $u_i$ and $u_{i+1}$ which are both on the endpoints and define a non exceptional distance, we have $u_{i+1}-u_i>\alpha$, since $u_{i+1}-u_i \neq \alpha$ (by the lemma's assumption).

On the other hand, if $u_i$ is  not  on an  endpoint, $u_{i-1}$ and $u_{i+1}$ are on endpoints, say $u_{i-1}=x_s(r)$ and $u_{i+1}=x_t(r)$. Also,  $u_i$ has color different from both $u_{i-1}$ and $u_{i+1}$.
Let $A=\{u_i: u_i \textrm{ is  not  on an  endpoint}, \; 2\leq i\leq p+q-1\}$. Note that for all $u_i\in A$, we have $u_{i}-u_{i-1}=\alpha$ (as done in Phase 2) and $u_{i+1}-u_{i}\geq \alpha$ (due to the separation constraint). By the lemma's assumption, if the distance between endpoints $x_j(r)$ and $x_k(r)$ is non exceptional, $u_{i+1}-u_{i-1}\neq 2\alpha$, i.e., $u_{i+1}-u_{i-1}> 2\alpha$. This implies $u_{i+1}-u_{i}> \alpha$. %%NK: it is clear. because $u_{i}-u_{i-1}=\alpha$.
Since $u_i$ is not on an endpoint,  we can shift $u_{i}$ toward $u_{i+1}$ infinitesimally such that $u_{i}-u_{i-1} >\alpha$ and we still have $u_{i+1}-u_{i} >\alpha$. Similarly, we perturb all $u_i\in A$ to have a new solution in which the distance between any two blue and red centers defining a non exceptional distance is strictly greater than $\alpha$. 

Now we use this property to find a solution $\bar{U}$ with a covering radius less than $r$.
We show that $r$ can be decreased without hitting feasibility constraints, i.e., none of the faces change, centers remain on their own faces (covering constraint) and the distance between any two consecutive red and blue centers are at least $\alpha$.  

For computing $\bar{U}$, let $0<\epsilon<r$ be a positive
real number to be fixed later. \remove{NK: confusing. (such that $\bar{U}$ is feasible).} For each $u_i\in U$, there are two cases:
\\
\noindent {\bf Case 1:} If $u_i$ is on an endpoint, say $x_{i,1}(r)$  (resp. $x_{i,2}(r)$),
 let $\bar{u_i}= x_{i,1}(r-\epsilon)$ (resp. $\bar{u_i}= x_{i,2}(r-\epsilon)$), i.e., it moves with the endpoint.
\\
\noindent {\bf Case 2:} If $u_i$ is not on an endpoint, $u_{i-1}$ and $u_{i+1}$ are on endpoints, $u_{i-1}=x_s(r)$ and $u_{i+1}=x_t(r)$, and also we have $u_i = u_{i-1} + \alpha$. If the distance between $x_s(r)$ and $x_t(r)$ is exceptional, let $\bar{u}_i=x_s(r-\epsilon) +\alpha$. Otherwise, (the distance of $x_s(r)$ and $x_t(r)$ is non exceptional), let $\bar{u}_i=u_i$.

We want to find an $\epsilon>0$ such
that $\bar{U}=\{\bar{u}_1,\bar{u}_2, \dots , \bar{u}_{p+q}\}$ is a
feasible solution with  radius of covering  $r-\epsilon$.
To this end,  we control the displacements of the centers in $\bar{U}$ compared to their initial locations in $U$, such that $\bar{U}$ with radius of covering $r-\epsilon$  is feasible.

Firstly, after decreasing $r$ to $r-\epsilon$, the relative
order of the endpoints of the intervals should not change, i.e., the
displacement of an endpoint of a face $F_i$ should be less than
$||F_i||/2$, where $||F_i||$ is the distance between the endpoints of
face $F_i$ that is not zero because of lemma's assumption. This can be controlled by shifting less than $\delta _1 /2$, where $0<\delta _1< \min_{1\leq i\leq 2n-1}\{||F_i||\}$. 

Secondly, for satisfying the covering constraint, the
centers belonging to $A$ should remain in their faces, i.e., for $u_i\in A$,
$x_{i,1}(r-\epsilon)<\bar{u}_i<x_{i,2}(r-\epsilon)$. Regarding  Case 1, the displacement of point  $x_{i,1}$  (resp. $x_{i,2}$) should be less
than $(u_i-x_{i,1}(r))/2$ (resp. ($x_{i,2}(r)-u_i$)/2). Let 
$0<\delta_2< \min_{\forall u_i\in A} \{ u_i-x_{i,1}(r), x_{i,2}(r)-u_i \}$.
Satisfying the covering constraint can be guaranteed by shifting less than $\delta _2 /2$.

Finally, for satisfying
the separation constraint, for each successive pair of blue and red centers, $\bar{u}_{i+1}$ and $\bar{u}_i$, we should have $\bar{u}_{i+1}-\bar{u}_i\geq \alpha$.
%/
If both of such points are on endpoints, i.e., $u_i=x_s(r)$ and $u_{i+1}=x_t(r)$, we need to have $|x_s(r-\epsilon)-x_t(r-\epsilon)|\geq \alpha$.  If the distance between $x_s(r)$ and $x_t(r)$ is exceptional with the value of $\alpha$, equality  $|x_s(r)-x_t(r)|= \alpha$ is always true for all values of $r$. Otherwise (the distance is non exceptional), the displacement
of these endpoints should be less than $(|x_s(r)-x_t(r)|-\alpha)/2$. For satisfying the separation constraint, the displacement should be restricted to $\delta_3$, $0<\delta_3<\min_{x_s(r),x_t(r)} \{ |x_t(r)-x_s(r)|-\alpha\}$ where
$u_i=x_s(r)$ and $u_{i+1}=x_t(r)$ have different colors and their distance is non exceptional. 
\\
If one of $u_i$ or $u_{i+1}$ is not on an endpoint, say $u_i$, then $u_{i-1}$ and $u_{i+1}$ are on the endpoints, i.e.,  $x_s(r)=u_{i-1}$ and $x_t(r)=u_{i+1}$. If the distance between $x_s(r)$ and $x_t(r)$ is exceptional with value of $2\alpha$, then $|x_s(r)-x_t(r)|=2\alpha$ for all values of $r$. By Case 2, $|\bar{u}_i-x_s(r-\epsilon)|=
|x_t(r-\epsilon)-\bar{u}_i|=\alpha$. Otherwise (the distance between $x_s(r)$ and $x_t(r)$ is non exceptional), we have $\bar{u}_i=u_i$. We should have $|x_t(r-\epsilon)-\bar{u}_{i}|\geq \alpha$ and $|\bar{u}_i-x_s(r-\epsilon)|\geq \alpha$, i.e., the displacement of $x_t(r)$ and $x_s(r)$ should be less than $|x_t(r)-u_{i}|-\alpha$ and $|u_i-x_s(r)|-\alpha$, respectively. 
%%NK : Shouldnt we also say the same about u_i - x_s(r)-> done by Homa
So let $0<\delta_4<\min_{\forall u_i \in A} \{ |u_{i+1}-u_i|-\alpha , |u_i-u_{i-1}|-\alpha \}$ and restrict the displacement to $\delta_4$  to satisfy the separability constraint.

Therefore, by choosing  numbers $\delta_1$, $\delta_2$, $\delta _3$, and $\delta_4$, and
$0<\delta < \min \{\delta_1, \delta_2, \delta_3,\delta _4\}/2$,
because of continuity of the movement of endpoints on line  $\Line$, we can obtain a positive $\epsilon$ such that the displacement of an endpoint becomes at most $\delta$ when the radius decreases to $r-\epsilon$.
Consequently, there exists $\epsilon>0$ such that the balls centered at points in $\bar{U}$ with covering radius $r-\epsilon$ are a feasible solution.
\end{proof}
By \lemref{candidate}, an optimal solution has at least a pair of two endpoints at a non exceptional distance of $0$, $\alpha$ or $2\alpha$.
Since the interval
endpoints $\lept_i(r)$ and $\rept_i(r)$ are given by $\lept_i(r) = p_{i1} - \sqrt{r^2 - \sum_{j=2}^d p_{ij}^2}$,
and $\rept_i(r) = p_{i1} + \sqrt{r^2 - \sum_{j=2}^d p_{ij}^2}$, a candidate set for the optimal radius can be computed by solving the following equations for all $1\leq i,k \leq n$ :
$$ \left( \pnt_{i1} \pm \sqrt{r^2 - \sum_{j=2}^d p_{ij}^2} \right) - \left( \pnt_{k1} \pm  \sqrt{r^2 - \sum_{j=2}^d p_{kj}^2} \right) =0 \text{ or } \alpha \text{ or } 2\alpha,$$ 
where the distance between $\pnt_i$ and $\pnt_k$ is non exceptional.
Note that at least one of those equalities holds true, and these equations provide a finite number of solutions.
By solving these equations, we obtain $O(n^2)$ candidates for the optimal radius. Thus we have,

\begin{lemma}
There is a set of $O(n^2)$ candidates for the optimal radius $r^*$, and this set can be constructed in $O(n^2)$ time.
\end{lemma}
We compute the candidates for $r^*$ and perform a binary search over them using the feasibility testing algorithm to obtain the optimal radius. We have the following theorem.
\begin{theorem}
The constrained $(n,p\lor q, \alpha)$ problem can be solved in $O(n^2+ T_{DP}(n,p,q)\log n) = O(n^3 p q \log n)$ time.
\end{theorem}

\newpage
\bibliography{bibliography}

\end{document}